\documentclass[final,1p,authoryear]{elsarticle}

\usepackage{amssymb}

\usepackage{graphicx}      % include this line if your document contains figures
\usepackage{natbib}        % required for bibliography
\usepackage{xcolor}
\usepackage{mathtools}
\usepackage{makecell}
\usepackage[utf8]{inputenc}
\usepackage{ACSD-CR}

\usepackage{xspace}
\usepackage{soul}
\usepackage{mathtools}
\usepackage{makecell}
\usepackage{nicefrac}
\usepackage{float}
\usepackage{wrapfig}
\newcommand{\ie}{ i.\,e.\xspace}
\newcommand{\eg}{e.\,g.\xspace}

\definecolor{dgreen}{rgb}{0.0, 0.5, 0.0}

\usepackage{textcomp}

\makeatletter
\newcommand{\subalign}[1]{%
	\vcenter{%
		\Let@ \restore@math@cr \default@tag
		\baselineskip\fontdimen10 \scriptfont\tw@
		\advance\baselineskip\fontdimen12 \scriptfont\tw@
		\lineskip\thr@@\fontdimen8 \scriptfont\thr@@
		\lineskiplimit\lineskip
		\ialign{\hfil$\m@th\scriptstyle##$&$\m@th\scriptstyle{}##$\crcr
			#1\crcr
		}%
	}
}

\journal{Control Engineering Practice}

\begin{document}
	
\begin{frontmatter}


\title{Experimental verification of an online traction parameter identification method} 

\author[TUC]{Alexander Kobelski} 
\ead{alexander.kobelski@etit.tu-chemnitz.de}
\author[TUC,Sk]{Pavel Osinenko} 
\ead{p.osinenko@skoltech.ru}
\author[TUC]{Stefan Streif \corref{CA}}
\ead{stefan.streif@etit.tu-chemnitz.de}

\cortext[CA]{Corresponding author}
\address[TUC]{Technische Universität Chemnitz, Automatic Control and System Dynamics Laboratory, 
   Germany}
   
\address[Sk]{Skolkovo Institute of Science and Technology, Computational and Data Science and Engineering Center, 
   Moscow, Russia}  
 
\SETCR{\CRELSE{\\https://creativecommons.org/licenses/by-nc-nd/4.0\\https://doi.org/10.1016/j.conengprac.2021.104837}{Control Engineering Practice}{Alexander Kobelski, Pavel Osinenko, Stefan Streif}}

\begin{abstract}                % Abstract of not more than 250 words.
Traction parameters, that characterize the ground-wheel contact dynamics, are the central factor in the energy efficiency of vehicles.
To optimize fuel consumption, reduce wear of tires, increase productivity etc., knowledge of current traction parameters is unavoidable.
Unfortunately, these parameters are difficult to measure and require expensive force and torque sensors.
An alternative way is to use system identification to determine them. 
In this work, we validate such a method in field experiments with a mobile robot.
The method is based on an adaptive Kalman filter.
We show how it estimates the traction parameters online, during the motion on the field, and compare them to their values determined via a 6-directional force-torque sensor installed for verification.
Data of adhesion slip ratio curves is recorded and compared to curves from literature for additional validation of the method.
The results can establish a foundation for a number of optimal traction methods. 
\end{abstract}

%	\item Research highlight 1
%	\item Research highlight 2

\begin{keyword}
System identification \sep  Kalman filter \sep Vehicle Dynamics \sep Traction
\end{keyword}

\end{frontmatter}
%===============================================================================

\section{Introduction}

Tractors have a huge market share that is steadily growing throughout the world \citep{industry_report_tractors2020}.
Machines with less than 30~HP alone are expected to grow by 60.7~Billion dollars until 2025. 
However, fuel costs and environmental regulations put ever stronger requirements on productivity and energy efficiency of such heavy-duty vehicles.
Productivity and energy efficiency in turn are highly dependent on ground conditions, which are usually unknown and change dynamically.
Determining traction parameters requires knowledge of wheel-ground forces which can be measured by relatively expensive sensors.
Online identification is an alternative way and is investigated experimentally in this work.

\begin{figure}[h]
	\centering
	\includegraphics[width=0.97\linewidth]{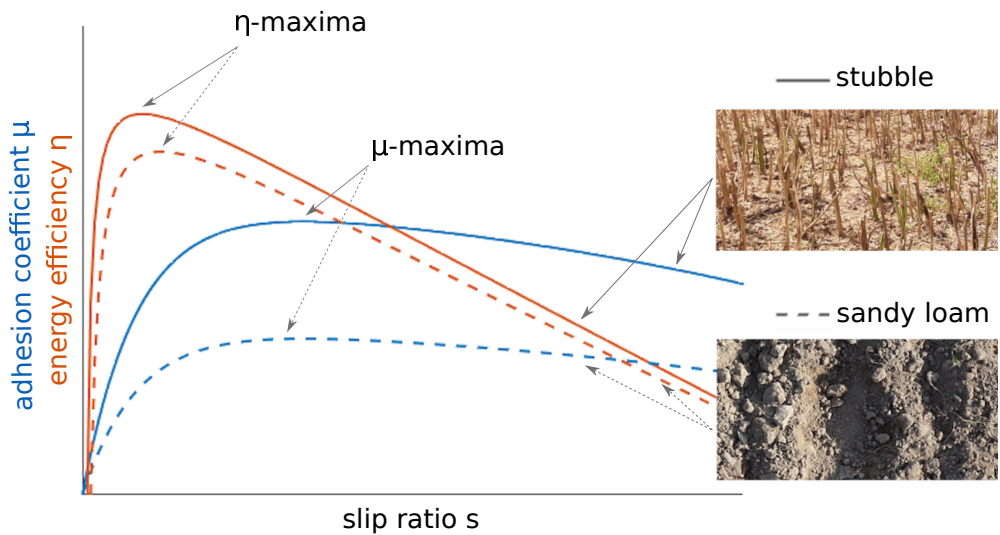}
	\caption{Typical $\mu(s)$ and  $\eta(s)$-characteristics, where $\mu$ refers to adhesion and $\eta$ to efficiency. Notice that both curves have their respective maximum at different slip ratios. The slip ratio for this maximum shifts with changing soil types and soil conditions, \eg moisture, tillage, respectively.}
	\label{fig:EE_AC_slip_graph}
\end{figure}

Energy efficiency characterizes the amount of driving power actually transmitted into pulling, whereas productivity is the ground speed times the working area.
Working area refers to width in case of tillage or depth times width in case of bulldozing.
The key traction parameters thereby are the adhesion and the rolling resistance coefficients.
They are the gross pulling force and, respectively, rolling resistance normalized by the vertical load.
Both are functions of the wheel slip ratio.
See Fig.~\ref{fig:EE_AC_slip_graph} for typical energy efficiency and adhesion coefficient depending on slip ratio.
The reader may refer to the classical work of \citep{Sohne1964} for a detailed description of traction slip ratio characteristics on different soils.
The overall operation performance can be understood in terms of balancing propulsion and energy efficiency and requires knowledge of the adhesion slip ratio curve $\mu(s)$ (see Fig.~\ref{fig:EE_AC_slip_graph}) as well as the rolling resistance coefficient.

Various models describing soil-wheel interaction have been developed, the approaches range from empirical to semi-empirical to finite element models.
Some are reviewed here.
\citet{Wismer1973} and \citet{Brixius1987} developed empirical models to predict traction performance using soil properties and tire parameters.
Their methods have found application in various works, \eg \citet{Kim2018} who used it in a slip controller.
\citet{Sandu_2011_SoftSoilTireModel} developed a semi-empirical off-road tire model for soft soil application which improves on existing models by providing a more detailed modeling of the soil-wheel contact surface.
\citet{Rajamani2012-adhesion-est} developed observers to estimate friction coefficients of individual wheels during operation from various measurements, \eg engine torque, brake torque and GPS measurements.
The topic keeps attracting attention of researches in recent time -- \citet{Pentos2017} used an artificial neural network to predict the influence of the soil texture, soil moisture, compaction etc. on the propulsion force and traction efficiency.
\citet{Addison2018} proposed using a data buffer coupled with a parameter fit algorithm for traction optimization.
\citet{Cook2019_TerrainCharacterization} recently developed a method for controlling tracked vehicles, where propulsion force, drive torque and slip ratio are estimated and compared to a list of known terrain traction models to find a suitable slip ratio set point.

This work will make use of Kalman filter based methods of traction identification.
\citet{Turnip2013-adhesion-est-EKF} used an identification approach based on the extended Kalman filter (EKF), whereas \citet{Hamann2014-adhesion-est-UKF} suggested to use a superior variant of the EKF -- the unscented Kalman filter (UKF), see \citep{Wan2000-UKF,VanDerMerwe2004-UKF}.
For an extensive review of wheel-soil-interaction models, traction parameter estimation, traction control schemes and wheel slip ratio estimation refer to \citep{SUNUSI_2020_Review_InteligentTractors}.
 
It is clear that the topic raised in this work is of high relevance and the reported results here contribute to the problem of online traction parameter identification.
In the following, we outline briefly the approach.

\begin{figure}[h]
	\centering
	\includegraphics[width=0.9\linewidth]{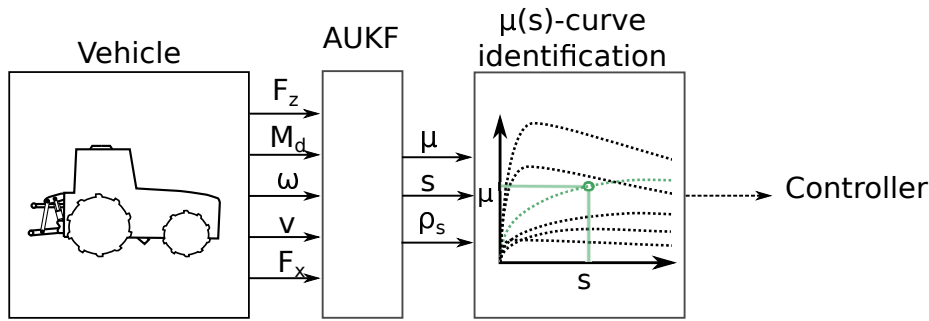}
	\caption{Diagram depicts the basic information flow of this works traction-identification algorithm. The vehicle provides sensor data to the adaptive unscented Kalman filter (AUKF), which estimates relevant traction variables. From single pairwise measurements of adhesion $\mu$ and slip ratio $s$ the algorithm computes whole adhesion slip ratio curves which can be used for control.}
	\label{fig:program_flow}
\end{figure}

First of all, in a previous work \citep{Kobelski_2020_TractionAndMapping} an UKF-based identification algorithm was used to estimate traction properties of a tractor driving over different ground types in a simulation.
It was based on the methodology developed in \citep{Osinenko2014,Osinenko2016,Osinenko2017}.
In \citep{Osinenko2016} the adaptive unscented Kalmanfilter with a fuzzy supervisor was used to assist traction control, while \citep{Osinenko2017} improved the model for the adhesion slip ratio characteristic curves.
This work follows up by validating the previously derived concept experimentally.
The goal of this work is to evaluate the tracking capabilities of the UKF under experimental conditions, record adhesion slip ratio curves for different soils and to test whether a change in ground conditions can be detected by the UKF.
A model for longitudinal motion dynamics based on dynamical equations combined with an adaptive UKF is used for the identification of the ground traction properties, namely, the adhesion coefficient $\mu$ and the soil deformation rolling resistance coefficient $\rho_{s}$. 
These two parameters are of the essence during slip control as they heavily influence productivity and energy efficiency, see Fig.~\ref{fig:EE_AC_slip_graph}.
Fig. \ref{fig:program_flow} shows a schematic of the overall algorithm.
The experimental design will be explained in detail, including the test vehicle with its sensor setup and the experimental site.
Statistical evaluation of the proposed algorithm is provided.

\section{Materials and Methods}
In this section, firstly, a model for the longitudinal motion dynamics of a vehicle derived from standard Newton equations is given.
An empirical model for adhesion slip ratio characteristics is introduced afterwards.
Then, the applied adaptive unscented Kalman filter is described.
In the last part, the design and setup of validation experiments is described.

\subsection{Motion dynamics}
\label{sec:traction_dynamics}

\begin{figure}[h]
	\centering
	\includegraphics[width=0.6\linewidth]{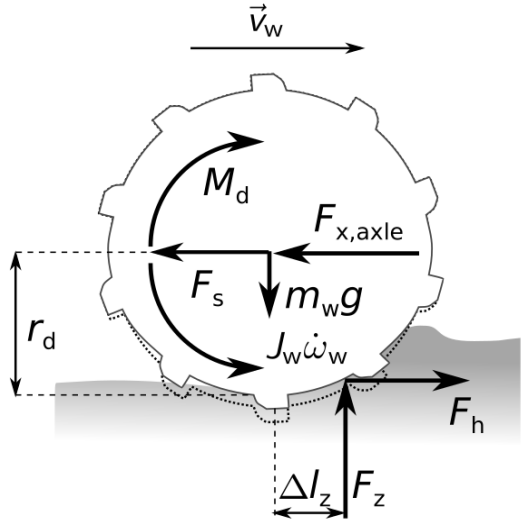}
	\caption{Diagram depicts forces and torque at a wheel. }
	\label{fig:wheel_forces}
\end{figure}

Model equations are derived in a similar fashion as in \cite{Kobelski_2020_TractionAndMapping}.
Consider the wheel force diagram of Fig.~\ref{fig:wheel_forces}.
There, the driving torque $M_\mathrm{d}$ exerts a horizontal force distribution in the contact area that can be summarized in a single force $F_\mathrm{h}$. 
The reaction of the vehicle's body $F_\mathrm{x,axle}$ is exerted in the opposite direction of $F_\mathrm{h}$. 
The wheel-ground contact results in the deformation of the tire and soil respectively, which in turn results in a tire deformation rolling resistance force $F_\mathrm{t}$ as well as a soil deformation rolling resistance force $F_\mathrm{s}$, with their respective rolling resistance coefficients $\rho_\mathrm{t}$ and $\rho_\mathrm{s}$.
The bearing friction is assumed proportional to the revolution speed $\omega$ with the resistance coefficient $\rho_\omega$.
The dynamical rolling radius $r_\mathrm{d}$ describes the deformed tire's distance between its center and ground.
The vertical (or normal) force $F_\mathrm{z}$ equals the sum of the axle load and wheel weight plus the vertical inertia force.
Due to the tire and soil deformation, the point of application of the soil reaction forces is shifted by $\Delta l_\mathrm{z}$.
By convention, the term $\Delta l_\mathrm{z}F_\mathrm{z}$ is assumed to be equal $r_\mathrm{d}F_\mathrm{t}$, where $F_\mathrm{t}=\rho_\mathrm{t}F_\mathrm{z}$, \ie the tire-deformation rolling resistance.
The force and and torque balance equations read:
\begin{align} 
m_\mathrm{w}\dot{v}_\mathrm{w} 			&= F_\mathrm{h} - F_\mathrm{x,axle} - F_\mathrm{s}, \label{eq:force_balance_horizontal} \\
J_\mathrm{w}\dot{\omega}_\mathrm{w} 	&= M_\mathrm{d} - r_\mathrm{d}F_\mathrm{h} - r_\mathrm{d}F_\mathrm{t}-r_\mathrm{d}\rho_\omega\omega_{w}. \label{eq:torque_balance} 
\end{align}
Here, $m_\mathrm{w}$ is the wheel mass, $v_\mathrm{w}$ is the ground speed and $J_\mathrm{w}$ is the wheel moment of inertia.
The total rolling resistance off-road is dominated by the soil-deformation resistance $F_\mathrm{s}$ and significantly influences the energy efficiency.
The soil deformation rolling resistance is summarized into $F_\mathrm{s}=\sum_{i=1}^{4}F_{\mathrm{s}i}=\rho_\mathrm{s}mg$.
With (\refeq{eq:force_balance_horizontal}) the longitudinal dynamics for the whole vehicle now are
\begin{equation}
m\dot{v} = \sum_{i=1}^{4}F_{\mathrm{h}i} - F_\mathrm{dx} - \rho_\mathrm{s}mg ,\label{eq:vehicle_dynamics}
\end{equation}
where $m$ is the vehicle mass, $F_\mathrm{dx}$ is the horizontal part of the drawbar pull, \ie the implement resistance and $i$ is the wheel index.

Dividing the horizontal force by the vertical force yield the adhesion coefficient $\mu$
\begin{equation}\label{eq:mu_physical}
\mu=\frac{F_\mathrm{h}}{F_\mathrm{z}}
\end{equation}
The part of the horizontal force that actually drives the wheel forward is called net traction ratio $\kappa$
\begin{equation}\label{eq:mu_rho_kappa}
\kappa=\mu-\rho_\mathrm{s}.
\end{equation}
The following definition of slip ratio is used:
\begin{equation}
\begin{array}{cccc}
s= & 1-\frac{|v|}{r_{d}|\omega_{w}|}, & \text{if} & |v|\leq r_{d}|\omega_{w}|,\\
s= & -1+\frac{r_{d}|\omega_{w}|}{|v|}, & \text{if} & |v|>r_{d}|\omega_{w}|.
\end{array}\label{eq:slip}
\end{equation}
It ranges from -1 (locked wheel) to 1 (spinning on the spot). 
The energy efficiency $\eta$ is defined by the formula:
\begin{equation}
\eta=\frac{\kappa}{\kappa+\rho}(1-s).\label{eq:traction-eff}
\end{equation}
Note that $\kappa$, $\mu$ and $\eta$ are functions of the slip ratio $s$.

The tire deformation resistance $\rho_\mathrm{t}$ mainly depends on tire type and inflation pressure and can therefore be estimated prior to operation, while the soil deformation resistance $\rho_\mathrm{s}$ and $\mu$ will be estimated using a state observer.
Due to vertical forces the wheel is deformed which reduces the rolling radius dynamically.
Please refer to \cite[p.40]{Guskov1988} for an empirical formula for estimation of the rolling radius $r_\mathrm{d}$.

The parameters  $J_\mathrm{w}$, $m_\mathrm{w}$, $m$ and $\rho_\omega$ are assumed known.
For details on estimation of $F_\mathrm{z}$ and $M_\mathrm{d}$, please refer to \citep{Osinenko2015a}.

\subsection{Adhesion Slip Ratio Characteristic Model}
\label{sec:adhesion-slip-model}
As was stated above, the adhesion coefficient is a function of the slip ratio (Fig.~\ref{fig:EE_AC_slip_graph}).
It was observed from literature, \eg \citet{Sohne1964}, that the shape of the adhesion-slip-curve ($\mu(s)$-curve) is  similar for similar ground types.

There exists a variety of $\mu(s)$-curve models, \eg \citep{Pacejka2006-veh-dyn,Schreiber2007} .
The empirical model from \citep{Schreiber2007} is used as a basis in this work.
The equation was first modified in \citep{Osinenko2014} by replacing a linear term with an exponential one und further refined in \citep{Osinenko2017}, by reducing the number of parameters by one:
\begin{equation}\label{eq:mu(s)_empirical}
\mu(s) = a(1 - p  e^{\alpha_1s} -  (1-p) e^{\alpha_2s}).
\end{equation}
Here, $a,p,\alpha_1,\alpha_2$ are the model parameters.
For similar ground types the three parameters $p,\alpha_1,\alpha_2$ may be fixed,  while only the parameter $a$ can be varied to cover a spectrum of ground conditions -- see Fig.~5 in \citep{Kobelski_2020_TractionAndMapping}.

\begin{figure}[!h]
    \centering
    \includegraphics[width=0.99\linewidth]{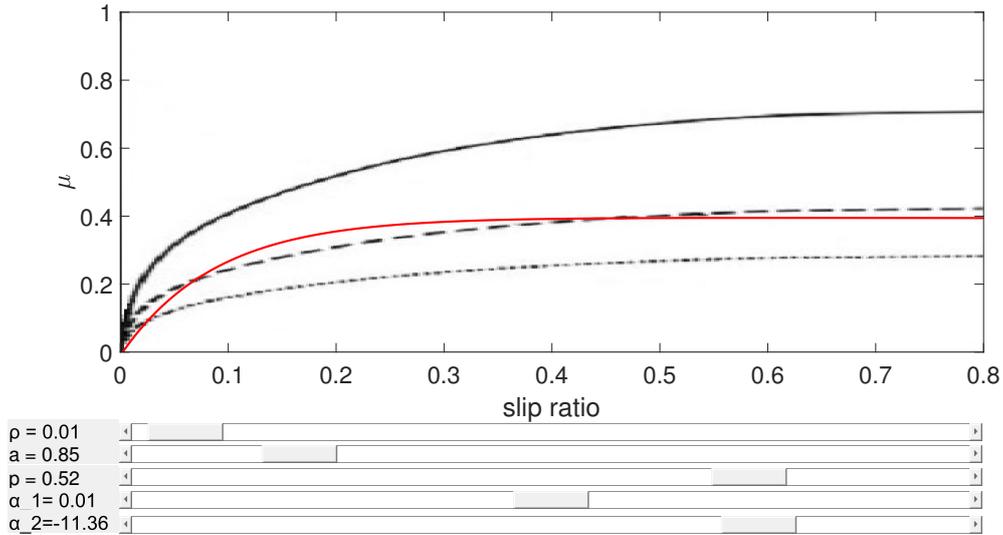}
    \caption{Figure shows functionality of the graphical fitting tool for the adhesion slip ratio curve. Shape of model curve (red) can be adjusted by moving the sliders at the bottom. The three black lines were added as a background picture with data from \citep{Wunsche2005}.}
    \label{fig:mu_tool}
\end{figure}

The $\mu(s)$-characteristics for each soil type can partially be obtained from literature, even before operation.
However, data recorded during operation may be more accurate due to varying local circumstances such as moisture, tilling or soil composition.
Hence, first a prototype $\mu(s)$-curve is fitted to data from literature using a graphical fitting tool, see Fig.~\ref{fig:mu_tool}.
This gives parameters $p,\alpha_1, \alpha_2$ from Eq.~\eqref{eq:mu(s)_empirical}.
The last parameter $a$ is then identified for every soil type encountered in the experiments separately through a least square (LS) fit to recorded data.
The identified curves are then evaluated statistically by comparing the identified $\mu(s)$-curves to recorded data, see section \ref{sec:EXP_soil_ident}. 
There the model choice of Eq.~\ref{eq:mu(s)_empirical} is also validated by experimental data.

Data from \citep{Wunsche2005} is used for the identification of the $\mu(s)$-prototype, see Fig.~\ref{fig:mu_tool}.
Fitting all four parameters of Eq.~\eqref{eq:mu(s)_empirical} is a non convex problem and will not result in a unique solution.
The graphical fitting tool allows the user to shape the prototype to fit well established data from literature or to self recorded data from experiments.
The downside, however, is that the shape forming is highly subjective to the user.
The next section explains, how the AUKF-FS identifies $\mu$, $\rho_\mathrm{s}$ and $s$.

\subsection{Adaptive Unscented Kalman filter}

The identification algorithm used in this work bases on the AUKF suggested in \citep{Jiang2007}.
The method has been throughly explained in previous work, while in this work only a basic understanding of its application is given (please refer to \citep{Osinenko2017} for a detailed explanation). 

The adhesion coefficient $\mu$ impacts vehicle performance the most and knowledge of it is required for efficient traction control.
For its identification, an adaptive UKF with a fuzzy-logic supervisor (AUKF-FS) is used.
Its purpose is to estimate the state $\mathbf{x}_{k}$ from the measured output $\mathbf{y}_{k}$. 
The generic model description behind the UKF reads: 
\begin{equation}
\begin{aligned} & \mathbf{x}_{k}=f\left(\mathbf{x}_{k-1},\mathbf{u}_{k-1}\right)+\mathbf{q}_{k-1},\\
& \mathbf{y}_{k}=h\left(\mathbf{x}_{k}\right)+\mathbf{r}_{k}.
\end{aligned}
\label{eq:discr-nonlin-model}
\end{equation}
Here, $\mathbf{x}_{k}\in\mathbb{R}^{n}$ is the state vector, $\mathbf{u}_{k}\in\mathbb{R}^{p}$ is the input vector, $\mathbf{y}_{k}\in\mathbb{R}^{m}$ is the output vector, $f\left(\mathbf{x}_{k-1},\mathbf{u}_{k-1}\right)$ is the non\textendash linear state model, $h(\mathbf{x}_{k})$ is the measurement model, $q_{k}\sim\mathcal{N}\left(0,\mathbf{Q}\right),r_{k}\sim\mathcal{N}\left(0,\mathbf{R}\right)$ are the state and measurement random noises with zero mean and covariance $\mathbf{Q}$ and $\mathbf{R}$ respectively, $\mathcal{N}$ denotes the normal probability distribution, $k$ is the time step index, $n,m,p\in\mathbb{N}$ are dimensions.
The AUKF algorithm can be divided into two steps: prediction and update.
The particular feature of the unscented Kalman filter is the calculation of the so called sigma-points for the prediction step.
The sigma-points are able to capture the posterior mean and covariance after propagation through the system up to the 3rd order of a Taylor series linearization (please refer to \citep{Wan2000-UKF} for details).
In the second part of the prediction the UKF computes the estimate probability distribution using the sigma-points as follows:

\begin{equation}
\begin{split}
&\text{PD}\left(\hat{\mathbf{x}}_{k|k-1}\big|\mathbf{y}_{1}...\mathbf{y}_{k-1}\right):=\\
&\mathit{\mathcal{N}}\left(\hat{\mathbf{x}}_{k|k-1}\bigg|\overset{2n}{\underset{i=0}{\sum}}\mathcal{W}_{m}^{(i)}\mathbf{\chi}_{k|k-1}^{(i)},\mathbf{P}_{k|k-1}\right).
\end{split}
\label{eq:UKF-predict-PDF}
\end{equation}

In \eqref{eq:UKF-predict-PDF}, $\mathbf{P}_{k|k-1}$ is the \emph{a priori} estimate covariance, and  $\chi_{k|k-1}^{(i)}=$$f\left(\chi_{k-1|k-1}^{(i)},\mathbf{u}_{k-1}\right)$ are the sigma\textendash points with the weights  $\mathcal{W}_{c}^{(i)}, \mathcal{W}_{m}^{(i)},i=0,...,2n$.
The predicted mean is computed from the \linebreak sigma\textendash points by the formula:

\[
\hat{\mathbf{x}}_{k|k-1}=\sum_{i=0}^{2n}\mathcal{W}_{m}^{(i)}\chi_{k|k-1}^{(i)}.
\]

The \emph{a priori} estimate covariance is calculated as follows:

\begin{equation}
\begin{split}
\mathbf{P}_{k|k-1}=\overset{2n}{\underset{i=0}{\sum}} & \mathcal{W}_{c}^{(i)}\left(\chi_{k|k-1}^{(i)}-\hat{\mathbf{x}}_{k|k-1}\right) \cdot \\
& \left(\chi_{k|k-1}^{(i)}-\hat{\mathbf{x}}_{k|k-1}\right)^\top+\mathbf{Q},
\end{split}
\label{eq:a-priori-est-cov}
\end{equation}

where $\mathcal{W}_{c}^{(i)},i=0,...,2n$ are weight factors.

The update step involves recalculating the sigma\textendash points from \linebreak $\mathit{\mathcal{N}}\left(\hat{\mathbf{x}}_{k|k-1}|\mathbf{P}_{k|k-1}\right)$.
The mean of the predicted output 
\begin{equation*}
\hat{\mathbf{y}}=\sum_{i=0}^{2n}=\mathcal{W}_{m}^{(i)}h\left(\chi_{k|k-1}^{(i)}\right)
\end{equation*}
and covariance 
\begin{equation*}
\begin{split}
\mathbf{S}_k=\sum_{i=0}^{2n}\mathcal{W}_{c}^{(i)} \left(h\left(\chi_{k|k-1}^{(i)}\right)-\hat{\mathbf{y}}_k\right) \cdot  \\ 
\left(h\left(\chi_{k|k-1}^{(i)}\right)-\hat{\mathbf{y}}_k\right)^\top+\mathbf{R},
\end{split}
\end{equation*}
as well as the state and output covariance,
\begin{equation*}
\mathbf{C}_k=\sum_{i=0}^{2n}\mathcal{W}_{c}^{(i)} \left(\chi_{k|k-1}^{(i)}-\hat{\mathbf{x}}_{k|k-1}^{(i)}\right) \cdot \left(h\left(\chi_{k|k-1}^{(i)}\right)-\hat{\mathbf{y}}_k\right)^\top
\end{equation*}
are then used to calculate the Kalman gain $\mathbf{K}_k=\mathbf{C}_k\mathbf{S}_k^{-1}$.
The last step is to update estimate mean with
\begin{equation}
\hat{\mathbf{x}}_{k|k}=\hat{\mathbf{x}}_{k|k-1}+\mathbf{K}_k(\mathbf{y}_k-\hat{\mathbf{y}}_k)
\end{equation} 
and the a posteriori covariance now becomes 
\begin{equation}
\mathbf{P}_{k|k}=\mathbf{P}_{k|k-1}-\mathbf{K}_k\mathbf{S}_k\mathbf{K}_k^\top.
\end{equation} 

The choice of noise covariance $\mathbf{Q}$ is crucial -- a poor choice of $\mathbf{Q}$ may result in divergence issues or noisy estimates \citep{Fitzgerald1971-KF-divergence}.
The standard UKF was hence modified by adding an adaption matrix $\mathbf{A}_{k}$ which replaces the state noise with $\mathbf{A}_{k}\mathbf{Q}$ (for an extensive description, please refer to \citep{Osinenko2014}).
The adaption matrix $\mathbf{A}_{k}$ minimizes the difference between true measurements and measurement estimate, which helps avoiding divergence issues, however the estimates may become too noisy.
To balance out these effects, it was suggested to introduce a fuzzy-logic system (FLS) to supervise the UKF.
The FLS captures intensity of vehicle dynamics using difference equations for measurements of $\omega_\mathrm{w}$ and $v$ over a moving window (please refer to \cite{Osinenko2016} for details).
If the vehicle enters a phase of intense dynamics, the FLS factor is set high which gives priority to the adaptation matrix and prevents divergence. 
At steady phases, the FLS is set smaller, which reduces estimation noise.
The resulting filter is called AUKF-FS (refer to \citep{Osinenko2017} for details).

From the vehicle dynamics model of Section \ref{sec:traction_dynamics} the state vector for the AUKF-FS consists of wheel speed $\omega_{w}$, vehicle speed $v$, adhesion coefficient $\mu$ and soil deformation rolling resistance $\rho_{s}$.
\[
\mathbf{x}=\left(\begin{array}{ccccc}
\omega_{w1},\dots\omega_{w4}, & v, & \mu_{1},\dots\mu_{4}, & \rho_{s}\end{array}\right)^\top.
\]
The wheel speeds and vehicle ground speed form the output vector:
\[
\mathbf{y}=\left(\begin{array}{cc}
\omega_{w1},\dots,\omega_{w4}, & v\end{array}\right)^\top.
\]
The input vector includes the drive torques, front vertical force $F_\mathrm{zf}$ and longitudinal component of the drawbar pull:
\[
\mathbf{u}=\left(\begin{array}{c}
M_{d1},\dots M_{d4},F_{zf},F_{dx}\end{array}\right)^\top.
\]
In \citep{Kobelski_2020_TractionAndMapping} observability of the overall system as well as possibilities to measure or estimate $M_{d}$, $F_{zf}$ and $F_{dx}$ are discussed.
The tire deformation rolling resistance coefficients $\rho_{t}$ are assumed as fixed parameters.
Propagation of the sigma-points through the system formed by Eq. (\refeq{eq:force_balance_horizontal}) and (\refeq{eq:torque_balance}) is performed using fourth-order Runga-Kutta method.
It is assumed that the unknown parameters $\mu_{1},\dots\mu_{4},\rho_{s},F_{zr}$ do not change during one integration step so that their dynamics are neglected.

\subsection{Experiment Design}
\label{sec:exp}
A mobile robot sensor platform was used to evaluate concepts of this work experimentally.
In this section first the mobile robot, tools and the experimental site will be described. 
Afterwards experiments will be explained.
\begin{figure}
	\centering
	\includegraphics[width=0.99\linewidth]{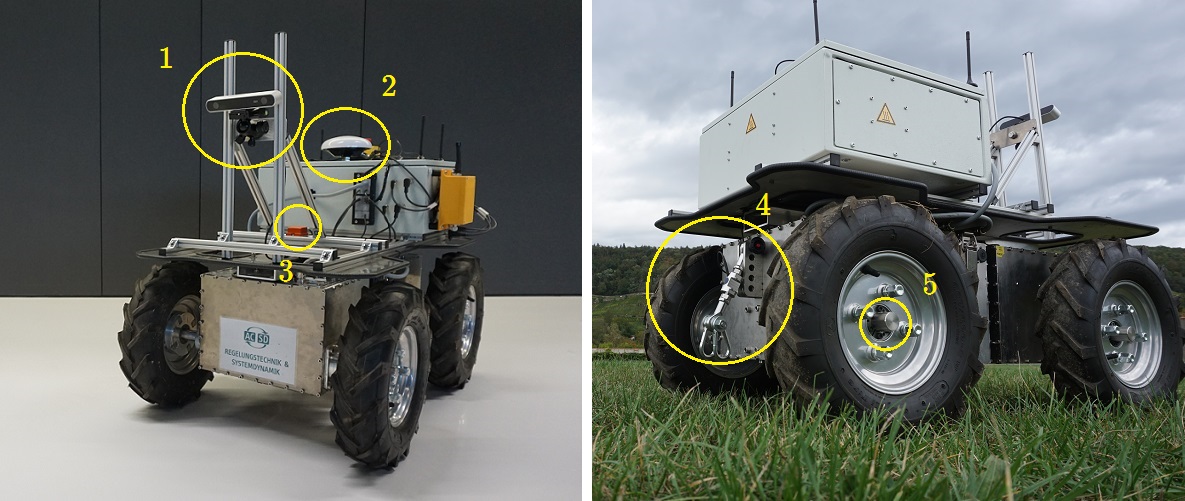}
	\caption{Mobile robot sensor platform from different perspectives. Position of some sensors is marked: 1 stereo camera, 2 GPS, 3 inertia measurement unit, 4 drawbar with force sensor, 5 force and torque sensor at wheel axis.}
	\label{fig:Pictures_Robot}
\end{figure}

\subsubsection{Tools and Experimental Site}
The mobile robot used is a customized version of the Innok Robotics Heros 444.
Its electric motors have a combined power of 1.6~kW, and each can be driven separately.
The wheels have a diameter of 40~cm and the vehicle weighs 139~kg, weight balance is 54~kg on the front and 85~kg on the rear axis.
It is equipped with a raspberry pi3 which functions as its brain, a NVIDIA Jetson TX2 for image processing and a LTE-WLAN-router.
It is equipped with a multitude of sensors:
\begin{itemize}
	\item Stereolabs ZED stereo camera for visual odometry 
	\item GPS Trimble BX982
	\item wheel odometry
	\item motor torque from electric motors
	\item inertia measurement unit XSens MTi-30-AHRS- 2A5G4
	\item force and torque sensor K6D80 at the back right wheel axis
	\item force sensor KM30z to measure drawbar pull force	
\end{itemize}
The robot is controlled with a remote controller during the experiments.
Wheel revolution speed can be adjusted with a rotatory switch.
In-built controllers ensure same revolution speed on all wheels.

\begin{figure}[h]
	\centering
	\includegraphics[width=0.95\linewidth]{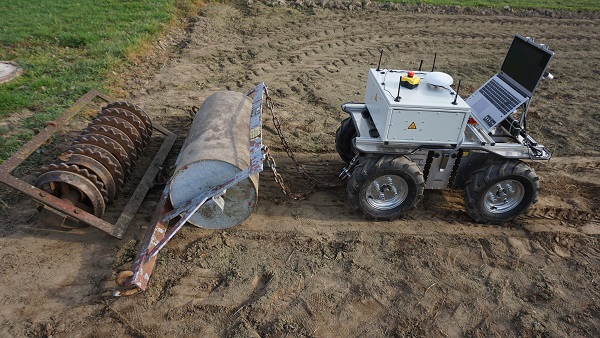}
	\caption{Left to right: A 'cultipacker'(Camebridge roller) is connected to a smooth surface roller which is connected to the mobile robot. Soil type is '\textbf{hard}'.}
	\label{fig:Vehicle_TwoTools_HardSoil}
\end{figure}

Various tools, see Fig.~\ref{fig:Vehicle_TwoTools_HardSoil}, were connected to the mobile robot's drawbar during the experiments.
Another tool, used in the experiments, is a heavy steel frame, similar to the one surrounding the smooth surface roller in Fig.~\ref{fig:Vehicle_TwoTools_HardSoil}.
Due to its structure, it could dig into the ground, which provided additional resistance during the experiments.
Pulling different tools across ground surfaces causes different horizontal forces.
Recall Eq.~\eqref{eq:mu_physical}, diverse horizontal forces translate to a wide range of operation points that can be recorded.

\begin{figure}[h]
	\centering
	\includegraphics[width=0.99\linewidth]{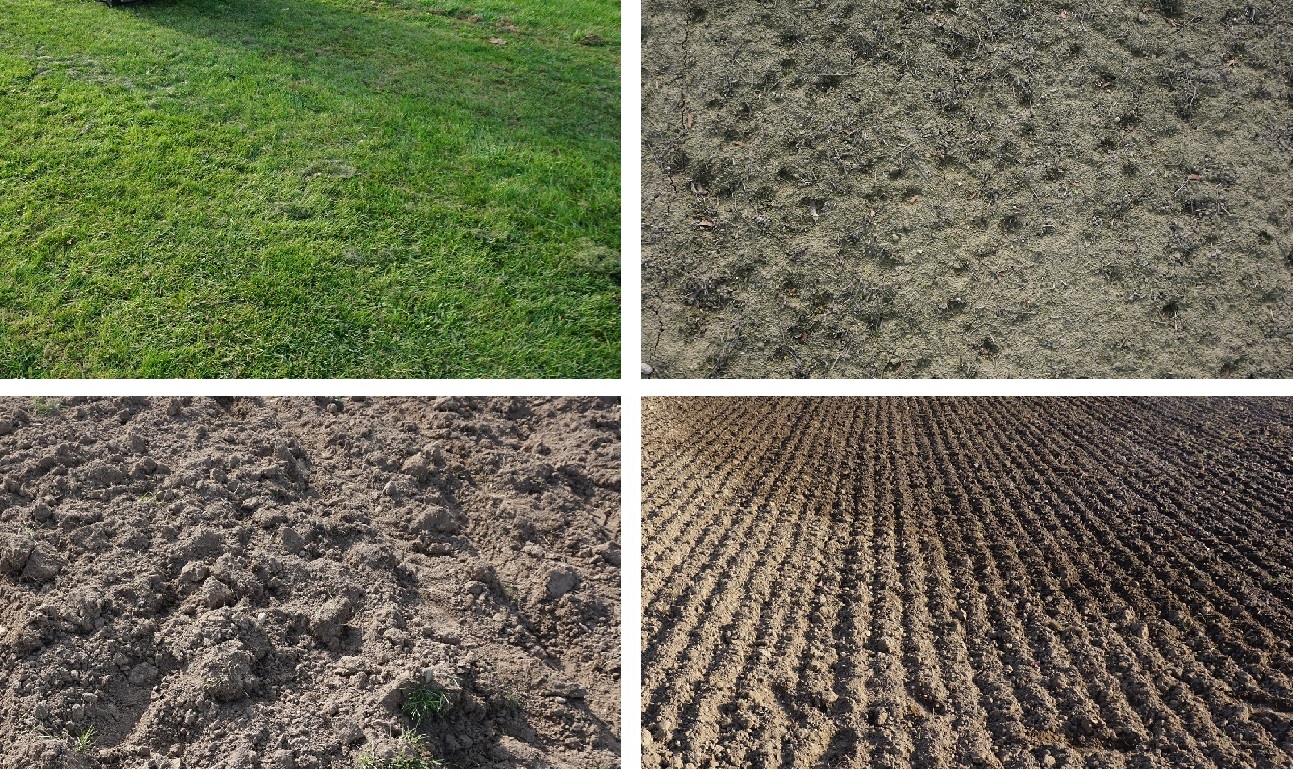}
	\caption{Ground types encountered during experiments. Top left: \textbf{grass}, top right: \textbf{hard} soil, bottom left: \textbf{coarse} soil, bottom right: \textbf{fin}e and\textbf{ we}t soil.}
	\label{fig:GroundTypes_All}
\end{figure}

Experiments were performed at a test site in Dresden, Germany.
The soil in this region is mostly a sandy loam with pseudogley (soil value 69).
The afore mentioned agricultural tools were used for experiments. 
Moisture and tillage of test sites could be adjusted to perform experiments on different ground types.
Two separate test sites were provided for experiments.
On the first field only one firmly pressed ground was present.
On the second field three ground types were prepared.
This is a list of soil types encountered in the experiments: 
\begin{itemize}
	\item[\textbf{hard}]  	firm soil which provides good traction, only on the first field
	\item[\textbf{fine}]  	tilled soil, relatively loose
	\item[\textbf{wet}]   	same soil as \textbf{fine}, but moisturized for three hours with a sprinkler
	\item[\textbf{coarse}] 	more coarse and uneven than \textbf{fine} with numerous earth clumps
	\item[\textbf{grass}]  	short grass (less than 5~cm) is growing around the fields
\end{itemize}
See Fig.~\ref{fig:GroundTypes_All} for pictures of the different ground types.

\subsubsection{Description of Experiment}
\begin{figure}[h]
 	\centering
 	\includegraphics[width=0.99\linewidth]{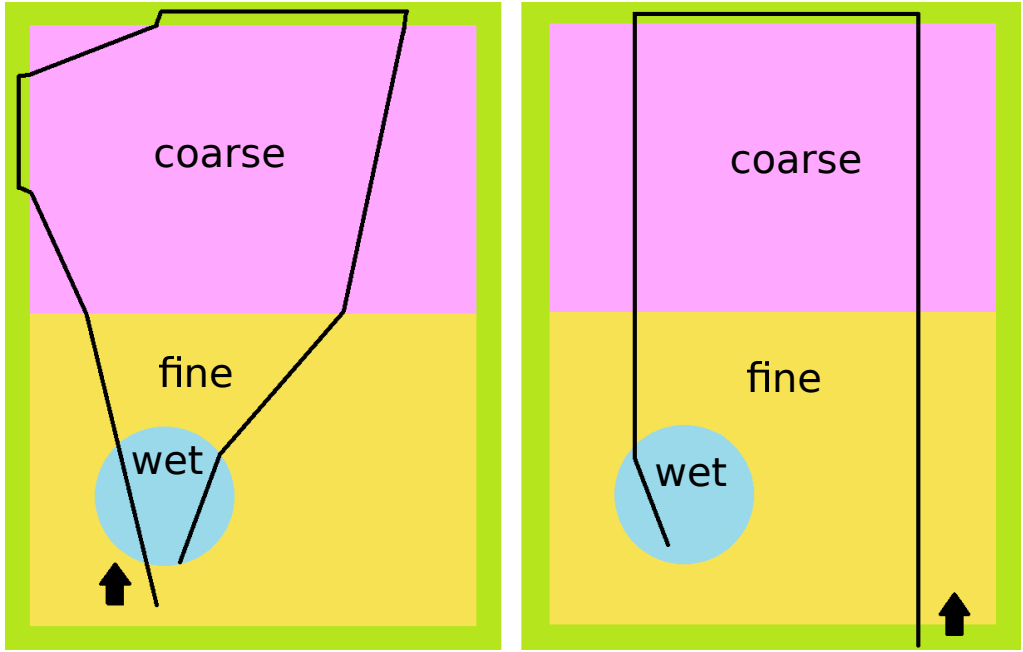}
 	\caption{Figure shows what path the mobile robot took during the two experiments '\textbf{multi1}'(left) and '\textbf{multi2}'(right). The arrow marks the start position and driving direction. The green border marks the grass section outside of the field. Real life pictures of the soil types can be seen in Fig.~\ref{fig:GroundTypes_All}}
 	\label{fig:driving_scheme}
 \end{figure}

Experimental data were stored on a Raspberry Pi and simultaneously transmitted to a laptop where it could be monitored in a Matlab/Simulink environment.
Experiments were conducted by driving the mobile robot with one or more tools connected to the drawbar over the various soil types.
68 experiments were conducted in total. 
Some experiments were conducted by driving over the same soil type with different speeds while others involve driving over switching ground types.
Note that the transition between two ground types is only finished when both the mobile robot and the pulled tool(s) are on the next ground.
This has to be kept in mind for evaluation as data recorded during the transitioning period are not representing the respective ground type.
Different combinations of tools as well as different wheel revolution speeds were tried during experiments, to receive data at multiple operation points on various soils.
Two experiments - they will be referred to in the following as '\textbf{multi1}' and '\textbf{multi2}' - involve driving over multiple ground types during one experiment, see Fig.~\ref{fig:driving_scheme}.
These two particular experiments will be used to evaluate the UKF's ability to estimate $\mu$.
 
\section{Results and discussion} 
\label{sec:results}
The experiments had three main objectives.
The first objective was to test the UKF's ability to estimate $\mu$. 
The second was to record $\mu(s)$-characteristics for different soils.
The third objective was to determine whether soil types can be distinguished trough UKF estimates during operation.
This is a prerequisite for future classification during operation, \ie matching the underground to known ground types.

\subsection{UKF Traction Estimation}

The two experiments '\textbf{multi1}' and '\textbf{multi2}' will be used as a benchmark to evaluate the UKF's tracking ability.
The UKF estimate for $\mu$ will be henceforth called $\mu_\mathrm{UKF}$.
Adhesion coefficient $\mu$ is calculated through Eq.~\eqref{eq:mu_physical} and from measurements of $F_\mathrm{x}$ and $F_\mathrm{z}$ at the rear right wheel.%\pagebreak

\begin{figure}[!h]
	\centering
	\includegraphics[width=0.99\linewidth]{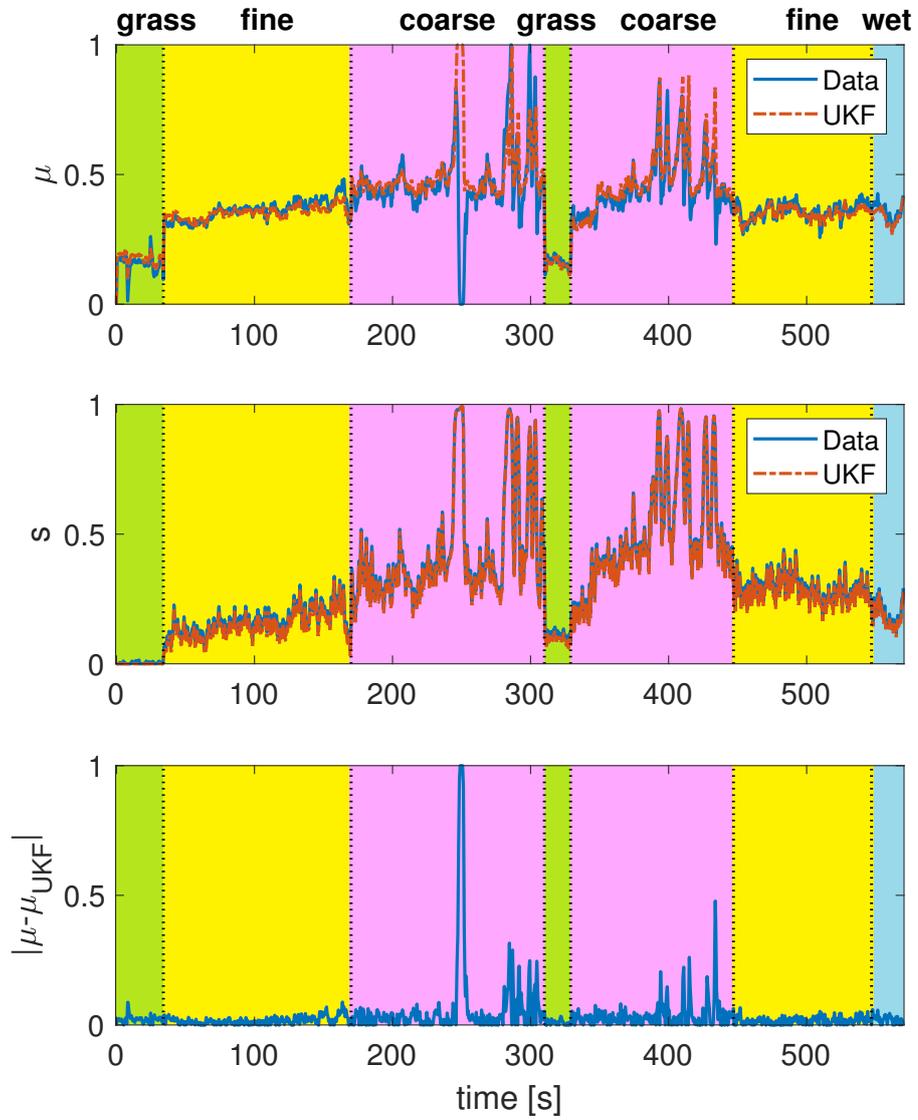}
	\caption{Comparison of measured $\mu$ versus estimated $\mu_\mathrm{UKF}$ during experiment '\textbf{multi2}',  $R^2=0.445$. Without the outlier between 245 to 255 seconds $R^2=0.848$. Fig.\ref{fig:driving_scheme} shows vehicle path during experiment while Fig.~\ref{fig:GroundTypes_All} shows real life pictures of the soil types.}
	\label{fig:mu_vs_MuUkf_f20}
\end{figure}

Fig.~\ref{fig:mu_vs_MuUkf_f20} shows recorded $\mu$ versus $\mu_\mathrm{UKF}$ during '\textbf{multi2}'.
At a few points $\mu_\mathrm{UKF}$ differs significantly from $\mu$.
At 250 seconds there is a wide gap between $\mu$ and $\mu_\mathrm{UKF}$, which is caused by the vehicle getting stuck in the '\textbf{coarse}' soil.
Due to its inertia, the tool connected to the vehicle would slide closer to the stuck vehicle.
The tool now pushes against the vehicle, resulting in negative or very small $F_\mathrm{x}$ and $\mu$, while motor torque -- which is the basis for the UKF estimate -- and $\mu_\mathrm{UKF}$ are high.
Similar phenomena occured around 280 to 310 seconds and 450 seconds. 
On coarse soil the robot would get stuck or drive over large earth clumps, resulting in uneven loads for the axes.
Data between 245 and 255 seconds are regarded as outliers and will be ignored during statistical evaluation.
Without outliers, $R^2$ is 0.85 and normalized root mean square error (NRMSE) is 0.09.
The root mean square error is normalized by the difference between maximum and minimum value of the observations.
The overall performance is comparable to earlier works, see \citep{Osinenko2014} and \citep{Osinenko2016}.

\subsection{Soil Characteristic Identification}
\label{sec:EXP_soil_ident}

\begin{figure}
	\centering
	\includegraphics[width=0.99\linewidth]{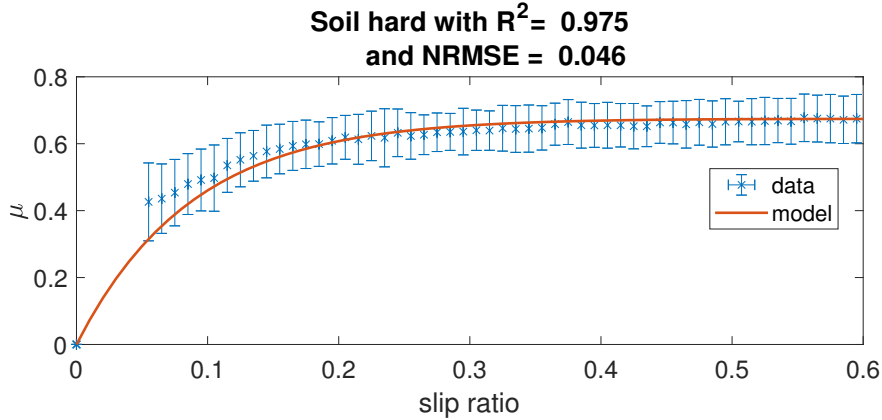}
	\caption{Figure shows data from experiments on 'hard' soil versus the model from Eq.~\eqref{eq:mu(s)_empirical}. Data are sorted by slip into 'bins', mean of $\mu$ for each respective bin is indicated by blue x, bars show standard deviation (SD) range (mean $\pm$ SD). Values for parameters $p,\alpha_1, \alpha_2$ were obtained with a graphic fitting tool. A least square fit was performed to find parameter $a$. Parameters are $a=1.42,\; p=0.52,\; \alpha_1=0.01,\; \alpha_2=-11.36$.}
	\label{fig:mu_s_fit_soil_hard}
\end{figure}

In section \ref{sec:adhesion-slip-model} a model, see Eq.~\eqref{eq:mu(s)_empirical}, for the adhesion slip ratio curve was given.
Using data from literature and a graphical fitting tool, a prototype curve was found.
In this section a scaling factor $a$ for the prototype curve will be fit to data from experiments.
This will validate the choice of the prototype curve and give different adhesion slip ratio curves for the soil types encoutered in the experiments.

\begin{table}
    \centering
    \begin{tabular}[c]{*{6}{p{1.8cm}}}
        \hline 
        &	\textbf{Hard}	& \textbf{Fine} & \textbf{Wet} 	& \textbf{Coarse} 	& \textbf{Grass} \\ 
        \hline 
        NRMSE 	& 0.046				& 0.093			& 0.138 		& 0.088				&  0.15\\ 
        \hline 
        R$^2$ 	& 0.975				& 0.9 			& 0.76			& 0.91				& 0.84 \\ 
        \hline 	
        $ a$	& 1.42				& 0.85 			& 0.83			& 0.91				& 0.4 \\ 
        \hline 
    \end{tabular} 
    \caption{Statistical evaluation of curve fitting the model from Eq.~\eqref{eq:mu(s)_empirical} to experimental data.} \label{tab:stats_boxplots_soil}
\end{table}

\begin{figure}
	\centering
	\includegraphics[width=0.99\linewidth]{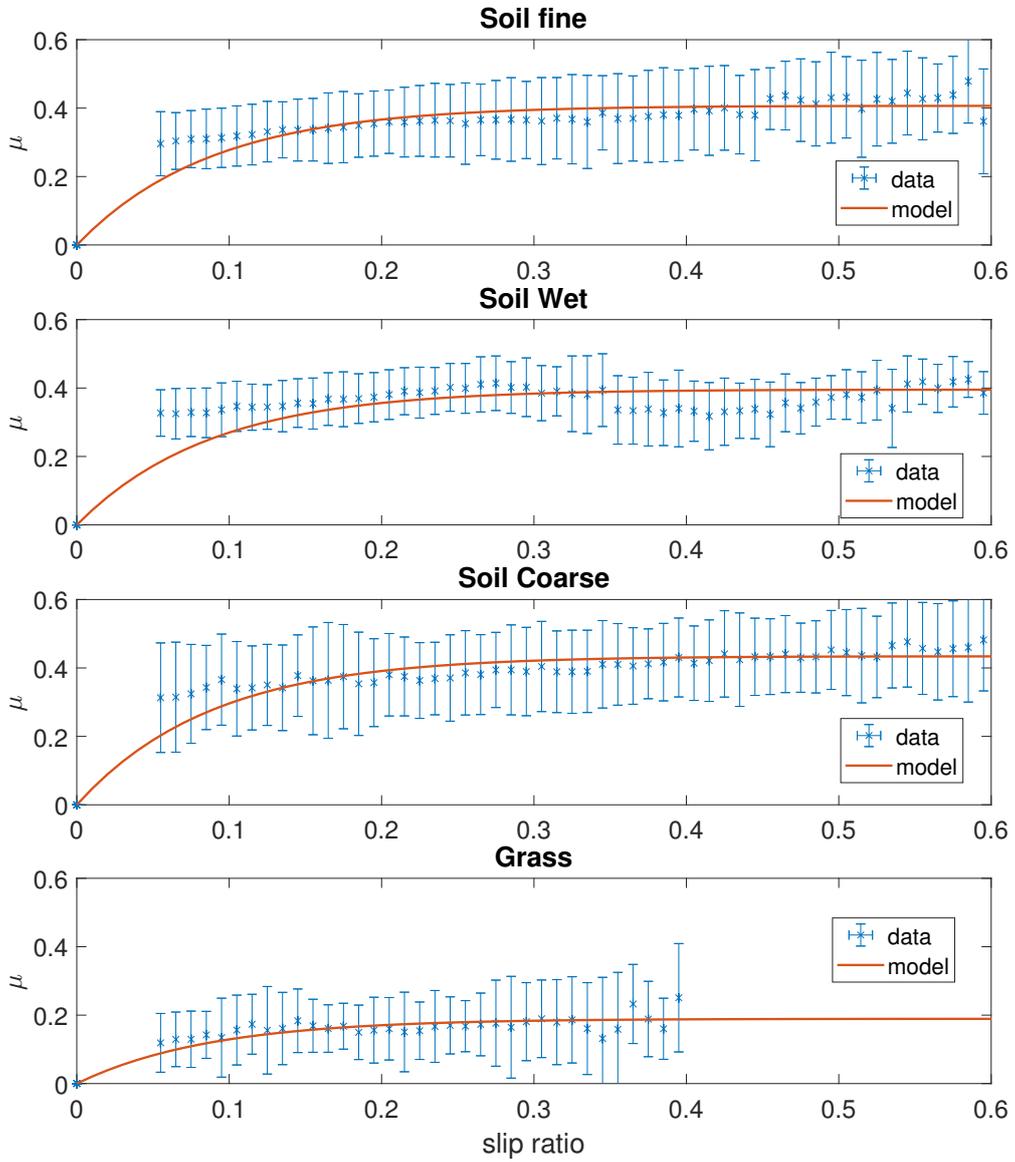}
	\caption{Figure shows data of adhesion $\mu$ and slip ratio $s$ on various soil types versus the model from Eq.~\eqref{eq:mu(s)_empirical}. Data are discretized and sorted by slip into 'bins', mean of $\mu$ for each respective bin is indicated by a blue x, bars show standard deviation (SD) range (mean $\pm$ SD). Values of parameter $a$ and quality of fit may be found in table \ref{tab:stats_boxplots_soil}. }
	\label{fig:musboxplots}
\end{figure}

From the experiments described in section \ref{sec:exp} five types of soil can be extracted, \ie~'hard', 'fine', 'coarse', 'wet' and 'grass'.
For the purpose of curve fitting recorded data are summarized in 'bins', where every 'bin' stores data from a certain slip range, \eg $5\%$ to $6\%$, $6\%$ to $7\%$ etc.
Data below 5\% and above 60\% slip ratio are neglected for fitting. 
For each bin, mean and standard deviation of $\mu$ are calculated, similar to a boxplot.
Curves are fit to the mean of each bin instead of the whole point cloud.
Fig. \ref{fig:mu_s_fit_soil_hard} shows the resulting fit for experiments on '\textbf{hard}'.
A visual comparison indicates a reasonably good fit, NRMSE and R$^2$ are 0.046 and 0.975 respectively.
Fig. \ref{fig:musboxplots} shows fitting for other soil types.
For the soil '\textbf{grass}' only values up to 0.4 slip ratio were used, measurements above 0.4 slip ratio were very few and were hence considered outliers.
Statistical evaluation is given in Table \ref{tab:stats_boxplots_soil}.

The soil types '\textbf{fine}' and '\textbf{wet}' can hardly be distinguished from each other, neither by comparing identified parameters ($a_\mathrm{fine}=0.85$ versus $a_\mathrm{wet}=0.83$) nor by comparing their standard deviations.
The identified parameter for 'coarse soil' is similar to the previous two ($a_\mathrm{coarse}=0.91$), however data show significantly higher standard deviation.
'\textbf{Grass}' shows overall a lower $\mu$ resulting in a lower parameter ($a_\mathrm{grass}=0.4$).
Furthermore, recorded data agree reasonably well with models from literature and with the model from Eq.~\eqref{eq:mu(s)_empirical}.

\subsection{Ground Change Detection}
This section examines whether differences in soil type can be observed from $s$ and $\mu$ during operation.
Data from experiments '\textbf{multi1}' and '\textbf{multi2}' is used as a benchmark for this. 
Recall that these were the two experiments, where the vehicle drove over multiple soil types.
Data are split into different sections, where cutting points are the transition zones between soil types, \eg from Fig.~\ref{fig:driving_scheme} it can be seen that '\textbf{multi1}' had three '\textbf{fine}' sections, three '\textbf{coarse}' sections etc.
For each section, where the vehicle drives over a single soil type, data are summarized in a mean and standard deviation value for $s$ and $\mu$ respectively.
Additionally UKF estimates are compared to measurements.

\begin{figure}
	\centering
    \includegraphics[width=0.9\linewidth]{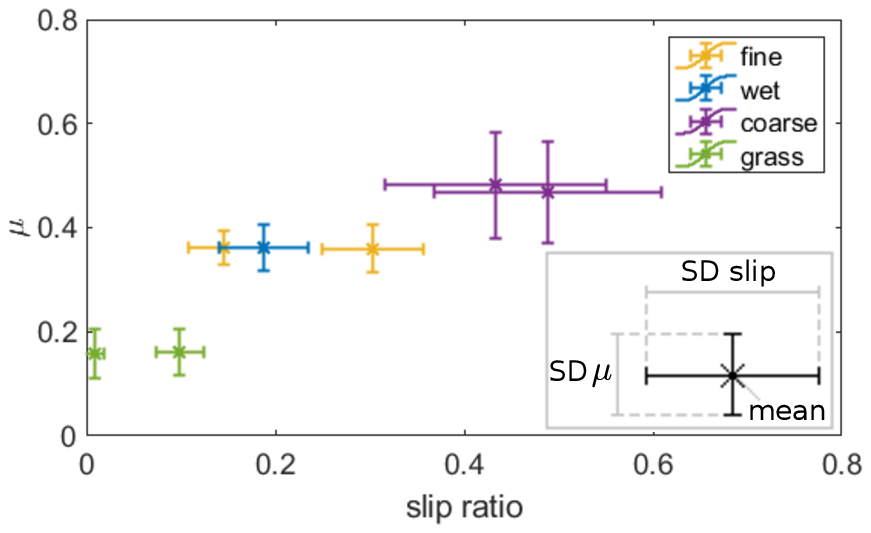}
	\caption{Figure shows mean and standard deviation for slip ratio $s$ and adhesion $\mu$ of individual sections during experiment '\textbf{multi2}'. Bars show standard deviation (SD) for $\mu$ and slip ratio respectively, center of each 'x' marks respective mean. Measurements on different soil types group up in their respective regions, showcasing that distinguishing between soils during operation is plausible.}
	\label{fig:mu_s_serrorbar_19plus20}
\end{figure}

\begin{figure}
	\centering
	\includegraphics[width=0.90\linewidth]{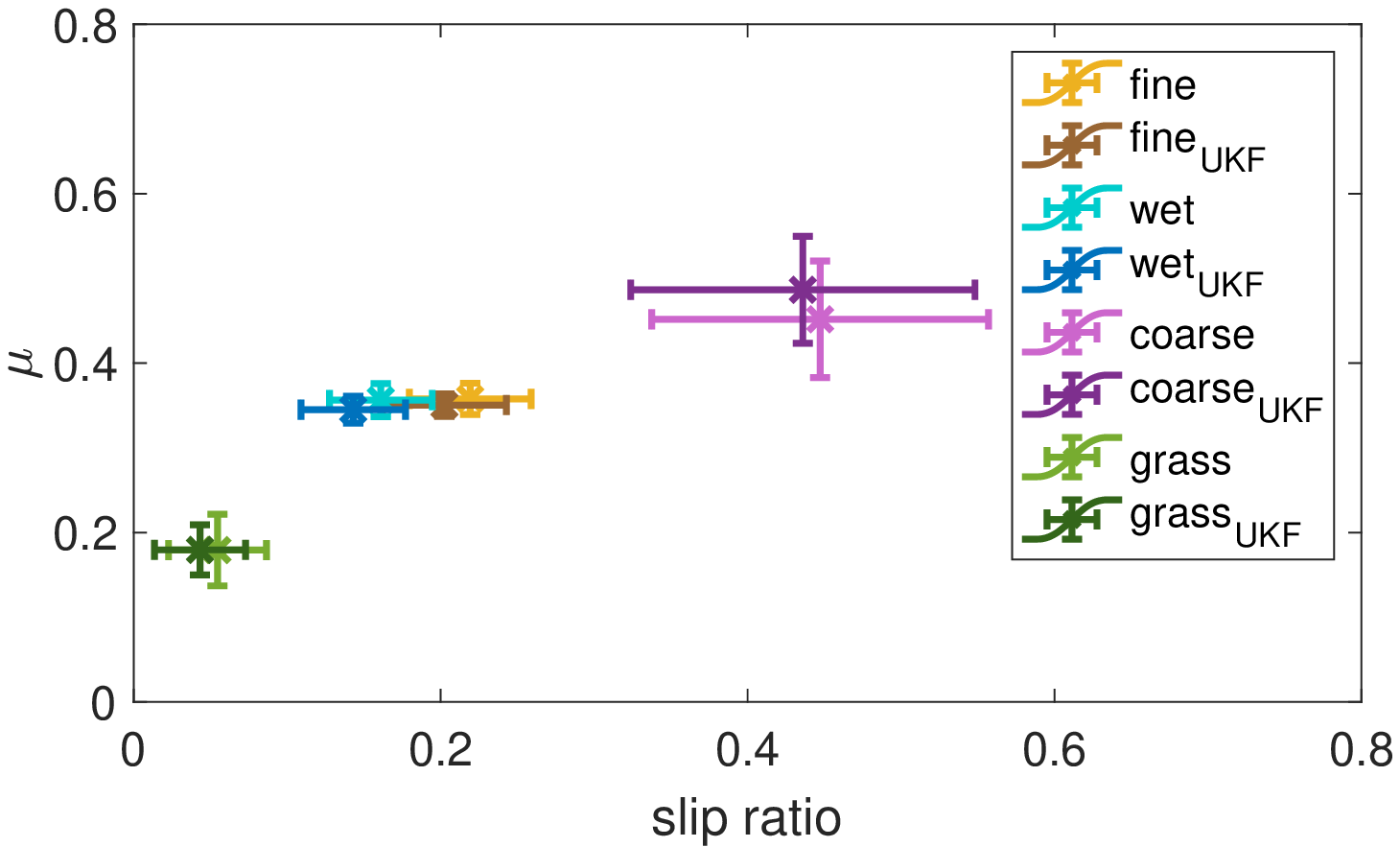}
	\caption{Figure is similar to Fig.~\ref{fig:mu_s_serrorbar_19plus20}, however data from '\textbf{multi1}' and '\textbf{multi2}' were used. Data for same soil types were combined. Additionally, UKF estimates of $s$ and $\mu$ are now depicted. Mean of $\mu$ and $\mu_\mathrm{UKF}$ for each soil type are close (less than 10~\% error) to each other, indicating a reasonably good accuracy.}
	\label{fig:mu_s_serrorbar_DataVsUKF}
\end{figure}

Fig.~\ref{fig:mu_s_serrorbar_19plus20} shows mean and standard deviation of $s$ and $\mu$ for various soil type sections during the two experiments.
Grass can easily be distinguished from the other soil types, as it has overall smaller $s$ and $\mu$.
On coarse soil the vehicle had slightly higher $\mu$ compared to fine and wet soil, however slip ratio mean and standard deviation are much higher.
Wet and fine soil can be distinguished from grass and coarse soil, however no difference between fine and wet can be reliably recognized from these experiments.
Perhaps the traction dynamics on these particular two soil types are too similar.

Fig.~\ref{fig:mu_s_serrorbar_DataVsUKF} compares mean and standard deviation from measurements to UKF estimates.
Differences between $\mu$ and $\mu_\mathrm{UKF}$ for '\textbf{grass}', '\textbf{wet}' and '\textbf{fine}' are smaller than 0.015, for '\textbf{coarse}' mean of $\mu$ and $\mu_\mathrm{UKF}$ difference is 0.03, \ie 8\% of $\mu$.
It can be concluded that a shift in operation point can be observed in the $\mu(s)$-characteristic, even from UKF estimates.
Therefore, distinguishing between soils during operation is plausible.

\section{Conclusion} \label{sec:conclusion}

Knowledge of traction parameters is essential when deriving control strategies for heavy-duty vehicles such as tractors.
Previous methods either involved offline estimation of said parameters, which has the downside of being less accurate due to dynamically changing conditions such as soil moisture or tilling, or required special and expensive measurements.
The method developed in this work does not require special sensors to work and adapts to changing conditions during operation.

Traction parameters are estimated using an adaptive unscented Kalman filter.
Then single measurements around a slip ratio operation point are used to estimate complete characteristic curves over a bandwidth of slip ratios.
These characteristic curves can then be used in control to derive strategies that balance energy efficiency, productivity and tire wearing.
In this work such a parameter estimation and curve identification algorithm was implemented on a mobile robot to verify the functionality in experiments.

Experiment evaluation shows reasonably good tracking capabilities of the AUKF estimation with $R^2=0.848$.
Earlier works, see \citep{Osinenko2014} and \citep{Osinenko2016}, showed similar results for their simulations.
Furthermore, identified adhesion slip ratio curves fit reasonably well to recorded experiment data, with $R^2$ ranging from $0.76$ to $0.975$.
In the last part of experiment evaluation it was tested, whether ground types are distinguishable during operation.
It showed, that indeed most soils can be distinguished from each other -- even with AUKF estimates -- with the exception of '\textbf{wet}' and '\textbf{fine}'.
The reason for that is most likely because compared to a tractor the mobile robot was much smaller and light weighted, hence its wheel would not dig as deep into the soil as a real tractor would. 
The '\textbf{wet}' soil hence did not impede vehicle movement as much as expected.
Overall it can be said, that a change of soil can be detected using this works AUKF estimation.
The results can be used in different ways.
The AUKF estimator is an easy and cheap to implement and works online during operation.
Knowledge of adhesion slip characteristics is essential for control and hence there is potential to improve energy efficiency and productivity of tractors.
Since operation points switch between soils with different traction properties, the underground may be identified and mapped using AUKF and Global Positioning System (GPS) information.
This can also be used to detect anomalies in the underground, \eg high $\mu$ may indicate firm soil which requires more tilling or patches of land where crops are growing more or less dense.

Future work will focus on the classification of soil types and ground recognition.
It will be tested whether the driving surface can reliably be identified using this works AUKF estimates.

\section*{Decleration of competing interest}
None declared.

\section*{Acknowledgments}

This research was founded by the Saxon Ministery of Science and Art and the 'Sächsische Aufbaubank (SAB)', SAB-project number 100333816.
For technical support we acknowledge the Saxon State Office for Environment, Agriculture and Geology.
We also would like to thank Christian Kröling for his excellent support in the realization of the field tests and the selection of the field equipment.

\bibliographystyle{elsearticle-harv}
\bibliography{bibtex/ctrl-general,bibtex/ctrl-misc,bibtex/electrical-engineering,bibtex/fuzzy-logic,bibtex/identification-misc,bibtex/Kalman-filter,bibtex/measurement,bibtex/motion-dynamics,bibtex/optimization,bibtex/signal-processing,bibtex/slip-ctrl,bibtex/traction-prediction,bibtex/tractors-misc,bibtex/vehicle-identification,bibtex/stochastics}

\begin{thebibliography}{26}
\expandafter\ifx\csname natexlab\endcsname\relax\def\natexlab#1{#1}\fi
\providecommand{\url}[1]{\texttt{#1}}
\providecommand{\href}[2]{#2}
\providecommand{\path}[1]{#1}
\providecommand{\DOIprefix}{doi:}
\providecommand{\ArXivprefix}{arXiv:}
\providecommand{\URLprefix}{URL: }
\providecommand{\Pubmedprefix}{pmid:}
\providecommand{\doi}[1]{\href{http://dx.doi.org/#1}{\path{#1}}}
\providecommand{\Pubmed}[1]{\href{pmid:#1}{\path{#1}}}
\providecommand{\bibinfo}[2]{#2}
\ifx\xfnm\relax \def\xfnm[#1]{\unskip,\space#1}\fi
%Type = Inproceedings
\bibitem[{Alexander et~al.(2018)Alexander, Sciancalepore and
  Vacca}]{Addison2018}
\bibinfo{author}{Alexander, A.}, \bibinfo{author}{Sciancalepore, A.},
  \bibinfo{author}{Vacca, A.}, \bibinfo{year}{2018}.
\newblock \bibinfo{title}{{Online Controller Setpoint Optimization for Traction
  Control Systems Applied to Construction Machinery}}, in:
  \bibinfo{booktitle}{Fluid Power Systems Technology}.
%Type = Article
\bibitem[{Brixius(1987)}]{Brixius1987}
\bibinfo{author}{Brixius, W.}, \bibinfo{year}{1987}.
\newblock \bibinfo{title}{{T}raction prediction equations for bias ply tires}.
\newblock \bibinfo{journal}{ASAE Paper} \bibinfo{volume}{87},
  \bibinfo{pages}{162}.
%Type = Article
\bibitem[{Cook et~al.(2019)Cook, Ray and
  Lever}]{Cook2019_TerrainCharacterization}
\bibinfo{author}{Cook, J.T.}, \bibinfo{author}{Ray, L.E.},
  \bibinfo{author}{Lever, J.H.}, \bibinfo{year}{2019}.
\newblock \bibinfo{title}{{Mobility Enhancement of Heavy-Duty Tracked Vehicles
  Under Load Using Real-Time Terrain Characterization, Traction Control, and a
  Towing Winch}}.
\newblock \bibinfo{journal}{Journal of Dynamic Systems, Measurement, and
  Control} \bibinfo{volume}{141}.
%Type = Article
\bibitem[{Fitzgerald(1971)}]{Fitzgerald1971-KF-divergence}
\bibinfo{author}{Fitzgerald, R.}, \bibinfo{year}{1971}.
\newblock \bibinfo{title}{{D}ivergence of the {Kalman} filter}.
\newblock \bibinfo{journal}{IEEE Transactions on Automatic Control}
  \bibinfo{volume}{16}, \bibinfo{pages}{736--747}.
%Type = Misc
\bibitem[{{Global Industry Analyst}(2019)}]{industry_report_tractors2020}
\bibinfo{author}{{Global Industry Analyst}}, \bibinfo{year}{2019}.
\newblock \bibinfo{title}{{G}lobal {F}arm {T}ractors {I}ndustry}.
%Type = Book
\bibitem[{Guskov et~al.(1988)Guskov, Velev, Atamanov, Bocharov, Ksenevich and
  Solonsky}]{Guskov1988}
\bibinfo{author}{Guskov, V.V.}, \bibinfo{author}{Velev, N.N.},
  \bibinfo{author}{Atamanov, Y.E.}, \bibinfo{author}{Bocharov, N.F.},
  \bibinfo{author}{Ksenevich, I.P.}, \bibinfo{author}{Solonsky, A.S.},
  \bibinfo{year}{1988}.
\newblock \bibinfo{title}{{T}raktory: {T}eoriya: {U}chebnik dlya {S}tudentov
  {V}uzov, po {S}pecialnosti "{A}vtomobili i {T}raktory". [{T}ractors.
  {T}heory. {T}extbook for {S}tudents of {H}igher {E}ducational {I}nsitutions
  {M}ajoring in {A}utomotive and {T}ractor {T}echnology (in {R}ussian)]}.
\newblock \bibinfo{publisher}{Moscow: Mashinostroenie}.
%Type = Inproceedings
\bibitem[{Hamann et~al.(2014)Hamann, Hedrick, Rhode and
  Gauterin}]{Hamann2014-adhesion-est-UKF}
\bibinfo{author}{Hamann, H.F.}, \bibinfo{author}{Hedrick, J.K.},
  \bibinfo{author}{Rhode, S.}, \bibinfo{author}{Gauterin, F.},
  \bibinfo{year}{2014}.
\newblock \bibinfo{title}{Tire force estimation for a passenger vehicle with
  the unscented kalman filter}, in: \bibinfo{booktitle}{Intelligent Vehicles
  Symposium Proceedings, 2014 IEEE}, \bibinfo{organization}{IEEE}. pp.
  \bibinfo{pages}{814--819}.
%Type = Article
\bibitem[{Kim and Lee(2018)}]{Kim2018}
\bibinfo{author}{Kim, J.}, \bibinfo{author}{Lee, J.}, \bibinfo{year}{2018}.
\newblock \bibinfo{title}{Traction-energy balancing adaptive control with slip
  optimization for wheeled robots on rough terrain}.
\newblock \bibinfo{journal}{Cognitive Systems Research} \bibinfo{volume}{49},
  \bibinfo{pages}{142 -- 156}.
%Type = Article
\bibitem[{Kobelski et~al.(2020)Kobelski, Osinenko and
  Streif}]{Kobelski_2020_TractionAndMapping}
\bibinfo{author}{Kobelski, A.}, \bibinfo{author}{Osinenko, P.},
  \bibinfo{author}{Streif, S.}, \bibinfo{year}{2020}.
\newblock \bibinfo{title}{Online traction parameter identification and
  mapping}.
\newblock \bibinfo{journal}{IFAC-PapersOnLine} \bibinfo{volume}{53}.
\newblock \bibinfo{note}{21st IFAC World Congress}.
%Type = Inproceedings
\bibitem[{Osinenko et~al.(2014)Osinenko, Geissler and
  Herlitzius}]{Osinenko2014}
\bibinfo{author}{Osinenko, P.}, \bibinfo{author}{Geissler, M.},
  \bibinfo{author}{Herlitzius, T.}, \bibinfo{year}{2014}.
\newblock \bibinfo{title}{Adaptive unscented kalman filter with a fuzzy
  supervisor for electrified drive train tractors}, in:
  \bibinfo{booktitle}{IEEE International Conference on Fuzzy Systems}.
%Type = Inproceedings
\bibitem[{Osinenko et~al.(2016)Osinenko, Geissler, Herlitzius and
  Streif}]{Osinenko2016}
\bibinfo{author}{Osinenko, P.}, \bibinfo{author}{Geissler, M.},
  \bibinfo{author}{Herlitzius, T.}, \bibinfo{author}{Streif, S.},
  \bibinfo{year}{2016}.
\newblock \bibinfo{title}{Experimental results of slip control with a
  fuzzy-logic-assisted unscented kalman filter for state estimation}, in:
  \bibinfo{booktitle}{2016 IEEE International Conference on Fuzzy Systems,
  FUZZ-IEEE 2016}, pp. \bibinfo{pages}{501--507}.
%Type = Article
\bibitem[{Osinenko and Streif(2017)}]{Osinenko2017}
\bibinfo{author}{Osinenko, P.}, \bibinfo{author}{Streif, S.},
  \bibinfo{year}{2017}.
\newblock \bibinfo{title}{{O}ptimal traction control for heavy-duty vehicles}.
\newblock \bibinfo{journal}{{C}ontrol Engineering Practice}
  \bibinfo{volume}{69}, \bibinfo{pages}{99 -- 111}.
%Type = Article
\bibitem[{Osinenko et~al.(2015)Osinenko, Geissler and
  Herlitzius}]{Osinenko2015a}
\bibinfo{author}{Osinenko, P.V.}, \bibinfo{author}{Geissler, M.},
  \bibinfo{author}{Herlitzius, T.}, \bibinfo{year}{2015}.
\newblock \bibinfo{title}{A method of optimal traction control for farm
  tractors with feedback of drive torque}.
\newblock \bibinfo{journal}{Biosystems Engineering} \bibinfo{volume}{129},
  \bibinfo{pages}{20 -- 33}.
%Type = Book
\bibitem[{{P}acejka(2006)}]{Pacejka2006-veh-dyn}
\bibinfo{author}{{P}acejka, H.B.}, \bibinfo{year}{2006}.
\newblock \bibinfo{title}{{T}yre and {V}ehicle {D}ynamics}.
\newblock Automotive engineering, \bibinfo{publisher}{Butterworth-Heinemann}.
%Type = Article
\bibitem[{Pentos and Pieczarka(2017)}]{Pentos2017}
\bibinfo{author}{Pentos, K.}, \bibinfo{author}{Pieczarka, K.},
  \bibinfo{year}{2017}.
\newblock \bibinfo{title}{Applying an artificial neural network approach to the
  analysis of tractive properties in changing soil conditions}.
\newblock \bibinfo{journal}{Soil and Tillage Research} \bibinfo{volume}{165},
  \bibinfo{pages}{113 -- 120}.
%Type = Article
\bibitem[{{R}ajamani et~al.(2012){R}ajamani, {P}hanomchoeng, {P}iyabongkarn and
  {L}ew}]{Rajamani2012-adhesion-est}
\bibinfo{author}{{R}ajamani, R.}, \bibinfo{author}{{P}hanomchoeng, G.},
  \bibinfo{author}{{P}iyabongkarn, D.}, \bibinfo{author}{{L}ew, J.Y.},
  \bibinfo{year}{2012}.
\newblock \bibinfo{title}{{A}lgorithms for {R}eal-{T}ime {E}stimation of
  {I}ndividual {W}heel {T}ire-{R}oad {F}riction {C}oefficients}.
\newblock \bibinfo{journal}{IEEE/ASME Trans. Mechatron.} \bibinfo{volume}{17},
  \bibinfo{pages}{1183--1195}.
%Type = Article
\bibitem[{Sandu et~al.(2011)Sandu, Pinto, Naranjo, Jayakumar, Andonian, Hubbell
  and Ross}]{Sandu_2011_SoftSoilTireModel}
\bibinfo{author}{Sandu, C.}, \bibinfo{author}{Pinto, E.},
  \bibinfo{author}{Naranjo, S.}, \bibinfo{author}{Jayakumar, P.},
  \bibinfo{author}{Andonian, A.}, \bibinfo{author}{Hubbell, D.},
  \bibinfo{author}{Ross, B.}, \bibinfo{year}{2011}.
\newblock \bibinfo{title}{Off-road soft soil tire model development and
  proposed experimental testing}.
\newblock \bibinfo{journal}{17th International Conference of the International
  Society for Terrain Vehicle Systems 2011, ISTVS 2011} ,
  \bibinfo{pages}{110--124}.
%Type = Article
\bibitem[{{S}chreiber and {K}utzbach(2007)}]{Schreiber2007}
\bibinfo{author}{{S}chreiber, M.}, \bibinfo{author}{{K}utzbach, H.},
  \bibinfo{year}{2007}.
\newblock \bibinfo{title}{{C}omparison of different zero-slip definitions and a
  proposal to standardize tire traction performance}.
\newblock \bibinfo{journal}{Journal of Terramechanics} \bibinfo{volume}{44},
  \bibinfo{pages}{75--79}.
%Type = Article
\bibitem[{S{\"o}hne(1964)}]{Sohne1964}
\bibinfo{author}{S{\"o}hne, W.}, \bibinfo{year}{1964}.
\newblock \bibinfo{title}{{A}llrad- oder {H}interradantrieb bei
  {A}ckerschleppern hoher {L}eistung [{A}ll whell or rear wheel drive train of
  farm tractors with high eingine power (in {G}erman)]}.
\newblock \bibinfo{journal}{Grundlagen der Landtechnik [Basics of agricultural
  engineering]} \bibinfo{volume}{20}, \bibinfo{pages}{44--52}.
%Type = Article
\bibitem[{Sunusi et~al.(2020)Sunusi, Zhou, Wang], Sun, Ibrahim], Opiyo,
  korohou, Soomro], Sale] and T.O.}]{SUNUSI_2020_Review_InteligentTractors}
\bibinfo{author}{Sunusi, I.I.}, \bibinfo{author}{Zhou, J.},
  \bibinfo{author}{Wang], Z.Z.}, \bibinfo{author}{Sun, C.},
  \bibinfo{author}{Ibrahim], I.E.}, \bibinfo{author}{Opiyo, S.},
  \bibinfo{author}{korohou, T.}, \bibinfo{author}{Soomro], S.A.},
  \bibinfo{author}{Sale], N.A.}, \bibinfo{author}{T.O., O.},
  \bibinfo{year}{2020}.
\newblock \bibinfo{title}{Intelligent tractors: Review of online traction
  control process}.
\newblock \bibinfo{journal}{Computers and Electronics in Agriculture}
  \bibinfo{volume}{170}, \bibinfo{pages}{105176}.
%Type = Article
\bibitem[{Turnip and Fakhrurroja(2013)}]{Turnip2013-adhesion-est-EKF}
\bibinfo{author}{Turnip, A.}, \bibinfo{author}{Fakhrurroja, H.},
  \bibinfo{year}{2013}.
\newblock \bibinfo{title}{Estimation of the wheel-ground contacttire forces
  using extended kalman filter}.
\newblock \bibinfo{journal}{International Journal of Instrumentation Science}
  \bibinfo{volume}{2}, \bibinfo{pages}{34--40}.
%Type = Inproceedings
\bibitem[{Van Der~Merwe et~al.(2004)Van Der~Merwe, Wan and
  Julier}]{VanDerMerwe2004-UKF}
\bibinfo{author}{Van Der~Merwe, R.}, \bibinfo{author}{Wan, E.A.},
  \bibinfo{author}{Julier, S.}, \bibinfo{year}{2004}.
\newblock \bibinfo{title}{{S}igma-point {K}alman filters for nonlinear
  estimation and sensor-fusion. {A}pplications to integrated navigation}, in:
  \bibinfo{booktitle}{Proceedings of the AIAA Guidance, Navigation \& Control
  Conference}, pp. \bibinfo{pages}{16--19}.
%Type = Inproceedings
\bibitem[{Wan and Van Der~Merwe(2000)}]{Wan2000-UKF}
\bibinfo{author}{Wan, E.A.}, \bibinfo{author}{Van Der~Merwe, R.},
  \bibinfo{year}{2000}.
\newblock \bibinfo{title}{{T}he unscented {K}alman filter for nonlinear
  estimation}, in: \bibinfo{booktitle}{Proceedings of the IEEE Symposium on
  Adaptive Systems for Signal Processing, Communications, and Control 2000.
  AS-SPCC}, pp. \bibinfo{pages}{153--158}.
%Type = Article
\bibitem[{Wismer and Luth(1973)}]{Wismer1973}
\bibinfo{author}{Wismer, R.}, \bibinfo{author}{Luth, H.}, \bibinfo{year}{1973}.
\newblock \bibinfo{title}{{O}ff-road traction prediction for wheeled vehicles}.
\newblock \bibinfo{journal}{Journal of Terramechanics} \bibinfo{volume}{10},
  \bibinfo{pages}{49--61}.
%Type = Book
\bibitem[{{W}{\"u}nsche(2005)}]{Wunsche2005}
\bibinfo{author}{{W}{\"u}nsche, M.}, \bibinfo{year}{2005}.
\newblock \bibinfo{title}{{E}lektrischer {E}inzelradantrieb f{\"u}r {T}raktoren
  [{E}lectrical single wheel drive for tractors (in {G}erman)]}.
\newblock Dresdner Forschungen / Maschinenwesen [Research in Dresden.
  Mechanical Engineering], \bibinfo{publisher}{TUDpress}.
%Type = Inproceedings
\bibitem[{{Zhe Jiang} et~al.(2007){Zhe Jiang}, {Qi Song}, {Yuqing He} and
  {Jianda Han}}]{Jiang2007}
\bibinfo{author}{{Zhe Jiang}}, \bibinfo{author}{{Qi Song}},
  \bibinfo{author}{{Yuqing He}}, \bibinfo{author}{{Jianda Han}},
  \bibinfo{year}{2007}.
\newblock \bibinfo{title}{A novel adaptive unscented kalman filter for
  nonlinear estimation}, in: \bibinfo{booktitle}{2007 46th IEEE Conference on
  Decision and Control}, pp. \bibinfo{pages}{4293--4298}.

\end{thebibliography}

\end{document}